\documentclass[11pt,twoside]{article}
\usepackage{./asp2014}

\aspSuppressVolSlug
\resetcounters
\def\xi{\hbox{$X_{\rm i}$}}
\catcode`\@=11
\def\gsim{\ifmmode{\mathrel{\mathpalette\@versim>}}
    \else{$\mathrel{\mathpalette\@versim>}$}\fi}
\def\lsim{\ifmmode{\mathrel{\mathpalette\@versim<}}
    \else{$\mathrel{\mathpalette\@versim<}$}\fi}
\def\@versim#1#2{\lower 2.9truept \vbox{\baselineskip 0pt \lineskip 
    0.5truept \ialign{$\m@th#1\hfil##\hfil$\crcr#2\crcr\sim\crcr}}}
\catcode`\@=12
\def\msun{\hbox{$M_\odot$}}
\def\msun{\hbox{$M_\odot$}}
\begin{document}

\title{Why Galaxies Care about AGB Stars: Setting the Stage}
\author{Alvio Renzini,$^1$ }
\affil{$^1$INAF - Osservatorio Astronomico di Padova, Padova, Italy \email{alvio.renzini@oapd.inaf.it}}

\paperauthor{Alvio Renzini}{alvio.renzini@oapd.inaf.it}{}{Istituto Nazionale di Astrofisica}{Osservatorio Astronomico di Padova}{Padova}{}{35121}{Italy}

\begin{abstract}
In this introduction to the Third Congress  of Vienna on AGB stars I first try to highlight why it is so hard to cope with the AGB evolutionary phase. This phase   is indeed dominated by three main  physical processes concerning bulk motions of matter inside/around stars, namely envelope convection, mixing and mass loss. They are inextricably  interlaced with each other in a circular sequence of reactions and counter-reactions which has so far undermined our attempts at calibrating such processes one independent of the other.  The second part of this introduction is focused on Globular Clusters, illustrating how they came to be a {\it new frontier} for the AGB evolution and a new opportunity to understand it.
\end{abstract}

\section{Why We Care}
We all know why galaxies care about AGB stars, but  it is perhaps worth listing a few items at the beginning of this III Congress of Vienna on AGB stars:

\begin{itemize}
\item {\bf The AGB Energetic role in Stellar Populations}, affecting the spectral energy distribution  (SED) of galaxies, especially in the near-IR, and consequently the mass and ages 
derived from SED fits.
\item {\bf The Chemical Yields} of several elements, notably He, Li, C, N, O, Ne, Na, Mg, Al, p-process and s-process elements, as they affect the chemical evolution of galaxies.
\item {\bf The Dust production}, with carbon-type  AGB stars being the source of carbonaceous grains, such as silicon carbide, sooth, graphite and perhaps even fullerene; and 
M-type AGB stars  being prime producers of silicate grains. Together, these dust grain obscure galaxies at low and high redshift.
\item  {\bf The Initial-Mass-Final-Mass Relation}, which  is  established during the AGB evolution, along with 
the mass return by dying stars in galaxies.
\end{itemize}

So, while we all agree that these are the issues at stake, the reason why we are all attending this conference is that AGB evolution is far from being properly understood.
In this introductory talk I will try to cursorily illustrate  what difficulties one encounters in modeling AGB stars, which explain why we still lack a satisfactory, quantitative understanding of this evolutionary phases, and therefore of the quantities for which galaxies care about.
\
\section{The Three Macrophysics Drivers of AGB Evolution}
The {\it input physics} we use to make stellar evolutionary models comes in two very distinct flavors: {\it microphysics} and {\it macrophysics}. The former deals with processes at the atomic and sub-atomic levels, which control three main ingredients in the models: equation of state, opacity, and nuclear reaction rates. Macrophysics instead deals with bulk mass motions inside/around stars, and include convection, rotation, mixing and mass loss. Virtually all the nasty problems in stellar evolution come from macrophysics and in this respect the AGB phase of intermediate mass stars is perhaps the nastier of all evolutionary phases. To simplify the discussion I will neglect rotation and magnetic fields, not because they cannot be important, but because considering just convection, mixing and mass loss makes life already quite complex, as I will try to illustrate. In the next subsections I will discuss them in turn, along with the interactions of each of them with the other two.

The reason why the mentioned macrophysics phenomena are a source of problems in modeling AGB stars is that for them there is no theory from first physical principles. What we are forced to do is to adopt for them some convenient, heuristic parametrization, and try to fix the free parameters by fitting observations. This is where all the problems arise.  The considerations that follow are most pertinent for AGB stars that experience the {\it hot bottom burning} (HBB) process, i.e., those with initial mass above $\sim 2.5-3\,\msun$, depending on composition, which are the most difficult to understand and where most of the action actually is taking place.

\subsection{Envelope Convection}
The envelope of AGB stars is convective and envelope convection is almost universally treated with the mixing-length theory (MLT), with the mixing length as a parameter. This parameter controls the stellar radius and could be fixed by matching stellar radii. However, the same parameter also controls the temperature at the base of the convective envelope: for example, in a $\sim 6\,\msun$ stellar model this temperature can go from $\sim 1$ to $\sim 100$ million K just by changing the mixing length by a factor of $\sim 2$. Thus, envelope convection controls the onset (and outcome) of the HBB process. In turn, the HBB controls the luminosity of the star, for given mass of its hydrogen-exhausted core. And this is not the end of the consequences, because luminosity and radius control mass loss and the radius affects also the amount of third dredge-up (TDU) at each thermal pulse. So, by just changing one parameter in the models, the mixing length, one affects radius, HBB, luminosity mass loss and TDU. All this might be fine, if we had a perfect theory of convection, but we know that the MLT is not, and therefore we have no assurance that  fixing the mixing length by matching one observable all other affected quantities will fall in the right place as well.

\subsection{Mixing}
The discovery by Icko Iben of the TDU has been one of the shining successes of the theory of stellar evolution, explaining the existence of carbon stars which had been a puzzle for over 100 years. Yet, the amount of dredged up material at each thermal pulse is extremely model dependent. This is traditionally indicated as $\lambda=\Delta M_{\rm dredge}/\Delta M_{\rm H}$, where
$\Delta M_{\rm dredge}$ is the mass of the dredged up layers after a thermal pulse and $\Delta M_{\rm H}$ is the advancement of the hydrogen burning shell during one interpulse period.
The net growth of the stellar core over a full thermal pulse cycle is therefore $(1-\lambda)\Delta M_{\rm H}$. Once more, the problem is that we do not have a theory of how mixing proceed at the edge between the fully convective envelope and the hydrogen-exhausted layers below it. In absence of a physical theory we have used plausible, yet heuristic arithmetical algorithms that may or may not capture reality.
Still, the amount of TDU, i.e., $\lambda$,  controls the evolution of the surface composition, the rate of core mass increase, hence the luminosity evolution which in turn controls mass loss while surface composition controls also radius and again mass loss. So, also the TDU, via $\lambda$, controls almost everything else, which in turn will control the TDU itself in a loop that nobody has really explored so far.

It is perhaps worth spending a few more words on the TDU-mass loss connection. For large values of $\lambda$  (near unity) the  core mass increases slowly, hence so does 
the luminosity,  implying  modest mass loss and  longer AGB lifetimes. With a small $\lambda$  (say, $\sim 0.1$) the  core mass increases rapidly and so does luminosity whose  increase  with core mass is extremely  steep when HBB is active (Bl\"ocker \& Sch\"omberner 1991). Thus, the star soon runs into dramatic mass loss leading to  a prompt abortion of the AGB phase. The result is a much different element yield  in the two cases, though the final mass of the WD remnant may not be so different after all. In other words, the HBB process works as an amplifier of the mixing-mass loss connection. Worth stressing is that we still have only {\it heuristic guesses} for the function $\lambda($mass, composition).

\subsection{Mass Loss}
Mass loss from red giants and supergiants, including AGB stars, is another fundamental process for which we don't have a physical theory. We even don't know if dust causes wind  or winds cause dust formation (personally, I incline for the second option). So, in AGB models mass loss is introduced via some simple formula giving the mass loss rate as a function of stellar mass, radius and luminosity, plus one adjustable parameter, for example what a long time ago I called $\eta$, for efficiency. Not to mention that on top of the regular wind we may also need a {\it superwind}, bringing in yet another parametrization and another parameter. So, mass loss, via these parameters, controls the mass evolution, hence the radius and the luminosity, which in turn affect the activity of the HBB process and the TDU. Also mass loss is therefore part of this inextricable network of feedbacks and counter-feedbacks.

\subsection{Wrup up}
Therefore, the  {\it Three Macrophysics Drivers } of AGB evolution are tightly interlaced to each other. This implies that one cannot calibrate each {\it free
parameter} independently of the others. Simple analytical formulations to
construct synthetic AGB evolutionary  sequences, or transplanted from one series of evolutionary models to another,  are bound to miss most of these intricate
connections. Only a tiny fraction of the AGB parameter space has been explored so far and the chance of having already found the right combination looks to me vanishingly small. AGB evolution is a particularly nasty puzzle! 

Still, the three macrophysics drivers do not complete the list of parametrized processes in constructing AGB evolutionary models. I have mentioned that rotation and magnetic fields are ignored here. Another process that can affect the AGB phase considerably is convective overshooting during the main sequence and pre-AGB evolutionary phases, which is  still poorly 
constrained by observations.

\section{Globular Clusters: A New Frontier for the AGB}
In recent years it became increasingly more likely that the study of globular clusters (GC) may offer an unexpected opportunity for a better understanding of AGB evolution, or at least for setting tighter constraints to it. It was once thought that GCs formed in a single burst, leading to prototypical simple stellar populations (SSP). On the contrary,  it now appears that  they host multiple stellar generations,
with secondary populations characterized by different chemical compositions and  our current  knowledge  can be summarized as follows:

\noindent
{\bf Ubiquity} Virtually all GCs that have been studied are composed of distinct stellar populations (at this moment
homogeneous data are available for 46 GCs, see Piotto et al. 2014 for a first release of HST Treasury
observations), from two in number up to six or more as in the case of NGC 2808 shown in Figure 1. Thus, the
production of MPs appears to be inextricably linked to the very formation of a GC.\\
{\bf Enrichment} The first-generation (1G) stars have a composition that can be ascribed to the proto-galactic matter
out of which they formed, whereas second-generation (2G) stars are depleted in C and O, enhanced in N and Na
and can be strongly enriched in helium (e.g., Norris 2004; Piotto et al. 2007; Carretta et al. 2009). This shows
that the material out of which they formed was exposed to proton captures at high temperatures (as it happens in AGB stars with HBB). With few
exceptions, 2G stars have the same abundances of iron and other heavy-elements as the 1G stars, indicating that
their material was not contaminated by supernova products. Even for GCs  showing a range of metallicities only a very small fraction of the 1G SN products has been incorporated in 2G stars. Huge cluster-to-cluster differences are also found in
the abundance of helium and p-capture elements of 2G stars.\\
{\bf Discreteness} One crucial property of MPs is that they split into distinct sequences in color-magnitude
diagrams and/or in two-color plots (see Figure 1),  as opposed to a continuous spread. This implies that 2G stars
formed in a sequence of discrete events, shortly after the formation of 1G stars. Figure 1 shows that UV and
UV-optical colors are especially effective in separating MPs, which is so because molecular bands, such as those of
OH, NH, CH and CN,  fall each in different UV/blue passbands, hence such colors depend on C:N:O proportions.

Figure 1 (from Milone et al. 2014, in preparation) offers a cogent example of these astonishing complexities for the case of the GC NGC 2808, clearly showing how the main sequence, the RGB and the HB split in distinct sequences and in the two color-color plots on the left side one can count at least five distinct populations. 

\articlefigure{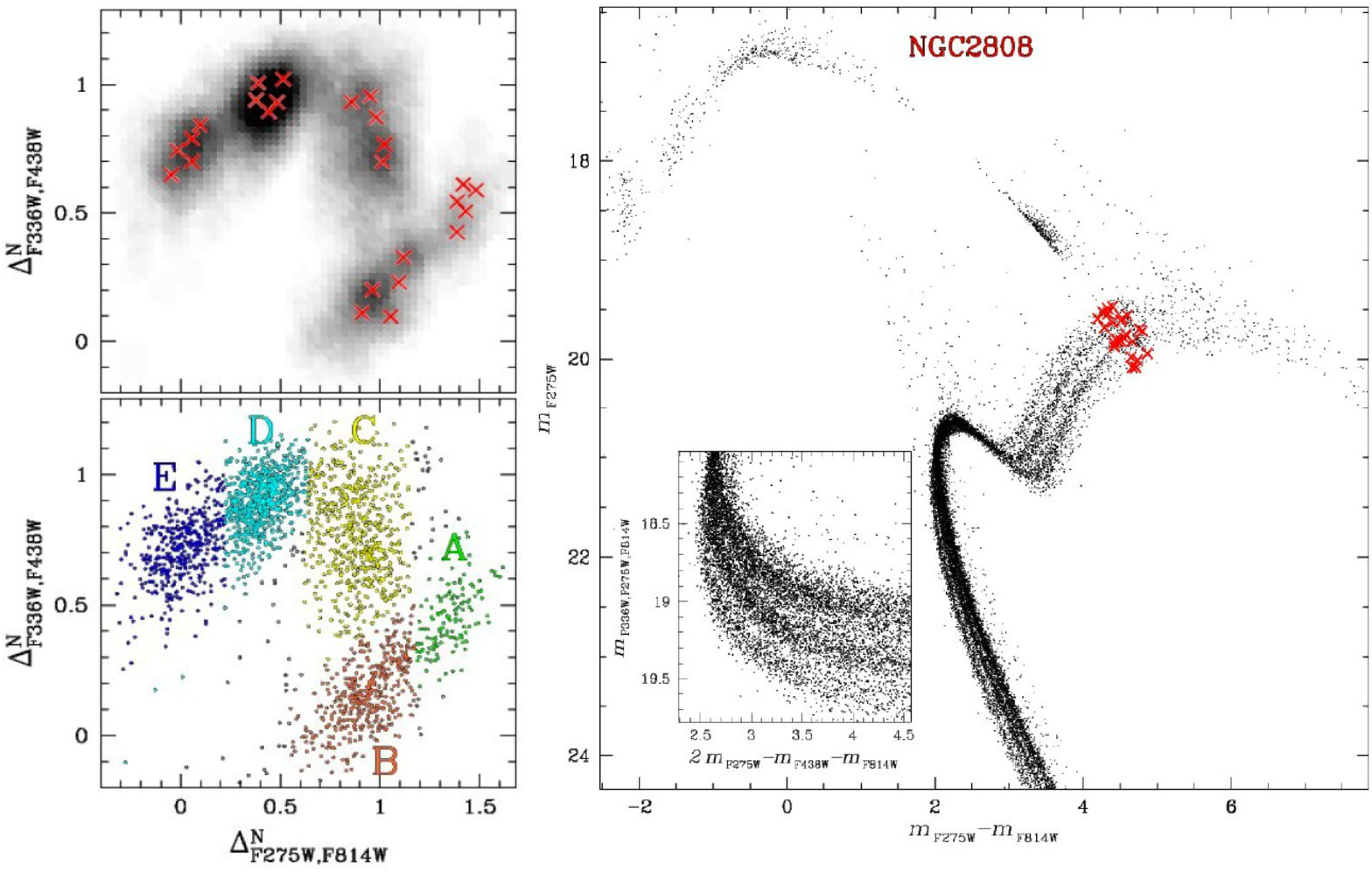}{ex_fig1}{Right panel: the HST color-magnitude diagram of the globular cluster  NGC~2808 combining the UV filter F275W of WFC3-UVIS with the near-IR F814W filter of ACS. Several discrete components are recognizable in all evolutionary sequences with the insert zooming on the main sequence turnoff and using a multicolor combination.  Left panels: a combination of four filters is used to split the red giant branch in at least five distinct stellar populations. Each clump corresponds to a population with a distinct pattern of He, C, N, O and Na abundances. Red crosses show stars being now observed with FLAMES at the VLT for their spectroscopic chemical characterization (from Milone et al. 2014, in preparation).}

Several  tentative progenitors of the 2Gs have been proposed. Massive stars (either fast rotating or in binary
systems) have been advocated for being the producers of the raw material out of which later generations formed,
but such scenarios have unsurmounting difficulties in explaining how to avoid supernova contamination and how to produce
quantized, distinct sub-populations (see Renzini 2008, 2013 for critical reviews). Thus, intermediate-mass stars
of the 1G in the mass range $3\lsim M\lsim 8-10\,\msun$  are widely regarded as the most plausible (or the least {\it im-plausible!})
candidates, with their accumulated AGB ejecta then experiencing a series of bursts of star formation. 

Also this scenario is facing two main difficulties. First, AGB models  do indeed produce helium enriched and oxygen-poor/sodium-rich, $p$-capture  material processed through the HBB, but existing AGB models do not quite reproduce the observed
O-Na anticorrelation. Second, only $\sim 5\%$ of the 1G mass can be returned to the ISM with the composition one
wants for the 2G stars. Since the 2G stars are often as numerous as the 1G stars (or even more so), the
immediate consequence is that the 1G ought to be at least $10-20$ times more massive than it is today (e.g.,
D'Ercole et al. 2008). Still, this assumes a ~100\% efficiency in the gas-to-star conversion, hence 1G had to be
even much more massive than that if the 2G star-formation efficiency was closer to a canonical ~10\%. This puts
the formation of GCs into a scale never conceived before: the progenitors of today's GCs with mass of a few
times 1$0^5$ to a few times $10^6\,\msun$, might have been stellar systems in the $10^7-10^8\,\msun$ mass range, or even more. With GCs being
the remnants of much bigger objects, the notion is also currently entertained that the majority of 1G stars, once tidally stripped off, made a major, if not dominant contribution to the formation of the Galactic halo. 

Thus, if multiple populations are  formed out of AGB ejecta, then the puzzling formation of GCs and AGB evolution become just one problem and the same.
Spectroscopic observations  are going to be used to further characterize each sub-population in terms of chemical composition, such as especially [O/Fe] and [Na/O] ratios. Main sequence splits will allow us the quantify differences in helium content as small as $\Delta Y=0.02$.  The full body of resulting compositions will then represent an {\bf extensive random sampling of AGB ejecta}, whose analysis will allow us to make  progress in understanding AGB evolution, a challenging but not impossible puzzle to assemble. Out of this it should be possible  to assess whether AGB models can be tuned to produce the observed composition patterns and, if so, one will have placed the tightest constraints on AGB evolution itself, including lifetimes and fuel consumptions as a function of stellar mass. This is indeed what is urgently needed to calibrate synthetic stellar population models  which are commonly used to estimate stellar masses and luminosity-weighted ages of distant galaxies, but where the AGB contribution to near-IR light is radically different from one set of models to another, leading to systematic differences which depend on age and are especially relevant for high redshift galaxies (see e.g., Pforr, Maraston \& Tonini 2012). 

\section{Conclusions}

So, many are the things we would like to know about  AGB Stars, such as  their yields of He, Li, C, N, O, Ne, Na, Mg, Al, s-process, the mass of hydrogen  and helium  burned during the AGB phase, hence the AGB contribution to the light (SED) of stellar populations, the time spent  and maximum  luminosity reached on AGB,  and all this as a function of initial stellar mass and composition. Stellar models can ``predict''  these quantities as functions of various, arbitrary input parameters, but almost certainly they are all  {\it wrong} as so far only a tiny fraction of the parameter space has been explored. ``All theories [of AGB evolution] are wrong, but --as one participants said at the meeting -- some theory is  more useful than others".  Mine indeed is not a plea against making AGB models. I would advocate that at this stage we may learn more on AGB evolution by {\it measuring} the just listed quantities on the multiple populations of globular clusters in the Milky Way and looking at the SED of  low- and high-redshift galaxies, especially those dominated by $\sim 1$ Gyr old stellar populations.
Then, by all means, we should  try to see if the models can be forced to do what we see.

\vskip 8mm
\noindent
{\bf Acknowledgements} I am grateful to the organizers for their invitation to attend this meeting which offered to me the opportunity to get exposed to the remarkable progress  achieved in recent years on the observations of AGB stars.


\end{document}